\documentclass[10pt]{IEEEtran}
\pdfoutput=1

\usepackage{subfigure}
\usepackage{float}
\usepackage{rotating}
\usepackage{wrapfig}
\usepackage{pgf,tikz}
\usetikzlibrary{arrows,automata,calc,shapes.arrows}
\usepackage{listings}
\usepackage{multirow}
\usepackage{wrapfig}
\usepackage{graphicx,color,xcolor}
\usepackage{xspace}
\usepackage{colortbl}
\usepackage{array}
\usepackage{amsmath,amssymb,amsthm,mathtools,stmaryrd}
\usepackage{macros}
\usepackage{myalg}
\usepackage{url}
\usepackage{booktabs}
\usepackage{etoolbox}
\usepackage[hidelinks]{hyperref}

\newtoggle{cameraready}
\togglefalse{cameraready}

\newcommand{\ghnote}[1]{\textbf{GH says: \textcolor{red}{#1}}}

\newcommand{\agnote}[1]{\textbf{AG says: \textcolor{red}{#1}}}

\title{Synthesizing Multiple Boolean Functions using
Interpolation on a Single Proof
 \thanks{This research was supported by the
  European Commission through project DIAMOND (FP7-2009-IST-4-248613),
  the Austrian Science Fund (FWF) through projects RiSE (S11406-N23) and
  QUAINT (I774-N23), and ERC Advanced Grant QUAREM (Quantitative Reactive Modeling).
\iftoggle{cameraready} {
}{
This paper originally appeared in FMCAD 2013, \url{http://www.cs.utexas.edu/users/hunt/FMCAD/FMCAD13/index.shtml}.
This version includes an appendix that is missing in the conference version.
}
  }
}

\author{Georg Hofferek${}^1$ \quad
        Ashutosh Gupta${}^2$ \quad
        Bettina K{\"o}nighofer${}^1$ \quad
        Jie-Hong Roland Jiang${}^3$ \quad
        Roderick Bloem${}^1$\\
${}^1$Graz University of Technology, Austria \qquad
        ${}^2$IST Austria \qquad
        ${}^3$National Taiwan University
      }

\begin{document}
 \maketitle

\begin{abstract}
It is often difficult to correctly implement a Boolean controller for a complex system, especially when concurrency is involved.  Yet, it may be easy to formally specify a controller.
 For instance, for a pipelined processor it suffices
to state that the visible behavior of the pipelined system should be
identical to a non-pipelined reference system (Burch-Dill paradigm).
We present a novel procedure to efficiently synthesize multiple
Boolean control signals from a specification given as a quantified
first-order formula (with a specific quantifier structure).  Our
approach uses uninterpreted functions to abstract  details of the
design. We construct an unsatisfiable SMT formula from the given
specification. Then, from just one proof of unsatisfiability, we use
a variant of Craig interpolation to compute multiple coordinated
interpolants that implement the Boolean control signals. Our method
avoids iterative learning and back-substitution of the control
functions.
We applied our approach to synthesize a controller for a simple
two-stage pipelined processor, and  present first experimental
results.

\end{abstract}

\section{Introduction}
\label{sec:introduction} Some program parts are easier to write than
others.  Freedom of deadlocks, for instance, is trivial to specify
but not to implement.  These parts lend themselves to synthesis, in
which a difficult part of the program is written automatically.  This
approach has been followed in program sketching
\cite{SketchingPLDI07,SketchingASPLOS06,SketchingPLDI05}, in lock
synthesis \cite{DBLP:conf/popl/VechevYY10}, and in synthesis using
templates
\cite{LoopFreeGulwaniPLDI11,InversionGulwaniPLDI11,SynthesisGulwaniPOPL10}.

In this paper, we consider systems that have multiple unimplemented
Boolean control signals. The systems that we will consider may not be
entirely Boolean.  We will consider systems with uninterpreted
functions, but our method extends to systems with linear arithmetic.
For example, consider a microprocessor.  Following Burch and Dill
\cite{Burch94}, we assume that a reference implementation of the
datapath is available.  Constructing a pipelined processor is not
trivial, as it involves implementing control logic signals that
control the hazards arising from concurrency in the pipeline.
Correctness of the pipelined processor is stated as equivalence with
the reference implementation. In this setting, we can avoid the
complexity of the datapath (which is the same in the two
implementations) by abstracting it away using uninterpreted
functions. Where Burch and Dill verify that the implementation of the
control signals is correct, we construct a correct  implementation
automatically.
This problem was previously addressed in \cite{Hoffer11}.  We improve
over that paper by directly encoding the problem into SMT, thus
avoiding BDDs, and by avoiding backsubstitution in case multiple
functions are synthesized.

Our approach is also applicable to synthesis of conditions in (loop-free)  programs.  As noted in~\cite{LoopFreeGulwaniPLDI11},
synthesizing loop-free programs  can be a
building block of full program synthesis. Prior work~\cite{SketchingPLDI07} presented various
techniques to deal with finite loops. Those techniques are also
applicable in our framework.

To synthesize a single missing signal, we can introduce a fresh
uninitialized Boolean variable $c$. We can express the specification
as a logical formula $\forall I \exists c \forall O.
\phiValid(I,c,O)$, which states that, for each input $I$, there
exists a value of $c$ such that each output $O$ of the function is
correct. Here, $I$ and $O$ can come from non-Boolean domains. If an
implementation is possible, the formula is valid and a witness
function for $c$ is an implementation of the missing signal.

Following~\cite{JiangICADD09}, we can generate a witness using
interpolation. In this paper, we generalize this approach by allowing
$n \geq 1$ missing components to be synthesized simultaneously. This
leads us to a formula of the form  $\forall I \exists c_1 \dots c_n
\forall O. \phiValid(I,c_1 \dots c_n,O)$. We use an SMT solver to
prove a related formula unsatisfiable and use
interpolation~\cite{McMillanSplitProver} to obtain the desired
witness functions. The first contribution of this paper is to extend
prior work~\cite{JiangICADD09} beyond the propositional level, and
consider formulas expressed in the theory of uninterpreted functions
and equality. As a second contribution, we propose a new technique,
called \emph{$n$-interpolation}, which corresponds to simultaneously
computing $n$ \emph{coordinated} interpolants from just one proof of
unsatisfiability. Like the interpolation procedures of
\cite{KovacsPOPL12,KovacsCADE09}, we need a ``colorable'' proof,
which we produce by transforming a standard proof from an SMT solver.

Our algorithm avoids the iterative interpolant computation described
in \cite{JiangICADD09}, where interpolants are iteratively
substituted into the formula. As the iterative approach needs one SMT
solver call per witness function, and interpolants may grow
dramatically over the iterations, this computation may be costly and
may yield large interpolants.  A similar back-substitution method is
also used in \cite{DBLP:conf/date/BloemGJPPW07} for GR(1) synthesis
and in~\cite{SynthesisKuncakPLDI10} for functional synthesis.
Our new method requires the expansion of the the (Boolean)
existential quantifier, increasing the size of the formula
exponentially (w.r.t. the number of control signals). Note, however,
that previous approaches~\cite{JiangICADD09} have the same
limitation.

\section{Illustration}
\label{sec:illus}

\begin{wrapfigure}{r}{3.5cm}
  \centering
  \usepgflibrary{shapes.geometric}
  \begin{tikzpicture}[>=stealth',shorten >=1pt,auto,
    node distance=9mm,semithick,minimum width = 5mm,minimum height = 5mm]
    \node[trapezium,rotate=-90,draw] (m1) {$1 \quad 0$};
    \node[rectangle,draw,yshift=-3mm, left of = m1] (n1){$neg$};
    \node[right of=m1] (o1) {$o_1$};
    \node[xshift=-4mm, left of=n1] (i1) {$i_1$};
    \draw (n1.east) -- (m1.319);
    \draw (m1.north) -- (o1);
    \draw (n1.west) -- (i1);
    \draw ([xshift=-0.4cm]n1.west) |- (m1.230);
    \node[xshift=1.7mm,yshift=1.7mm,below of=m1] (c2){$c_1$};

    \node[trapezium,yshift=-10mm, below of= m1,rotate=-90,draw]
    (m2){$1 \quad 0$};
    \node[rectangle,draw,yshift=-3mm, left of = m2] (n2){$neg$};
    \node[right of=m2] (o2) {$o_2$};
    \node[xshift=-4mm, left of=n2] (i2) {$i_2$};
    \draw (n2.east) -- (m2.319);
    \draw (m2.north) -- (o2);
    \draw (n2.west) -- (i2);
    \draw ([xshift=-0.4cm]n2.west) |- (m2.230);
    \node[xshift=1.7mm,yshift=-1.7mm,above of=m2] (c2){$c_2$};

    \node[yshift=-9.5mm, xshift= -2mm, left of= m1, rectangle,draw]
    (m3){\begin{array}[t]{@{}c@{}}  ??\\ \end{array}};
    \draw (m1) |- (m3.20);
    \draw (m2) |- (m3.340);
    \draw ([xshift=-0.4cm]n1.west) |- (m3.150);
    \draw ([xshift=-0.4cm]n2.west) |- (m3.210);
  \end{tikzpicture} \\

  \caption{Example of controller synthesis.}
  \vspace{-0.5cm}
  \label{fig:ex-hardware}
\end{wrapfigure}
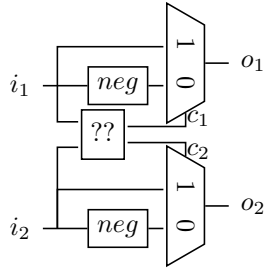

In this Section we illustrate our approach using a simple controller
synthesis problem. Figure~\ref{fig:ex-hardware} shows an incomplete
hardware design. There are two input bit-vectors $i_1$ and $i_2$,
carrying non-zero signed integers, and also two output bit-vectors
$o_1$ and $o_2$ carrying signed integers. The block $neg$ flips the
sign of its input. The outputs are controlled by two bits, $c_1$ and
$c_2$. The controller of $c_1$ and $c_2$ is not implemented. Suppose
the specification of the incomplete design states that the signs of
the two outputs must be different. Formally, the specification is
\begin{multline*}
\hspace{-2ex}\forall i_1 , i_2.\exists c_1, c_2. \forall o_1, o_2.
((c_1 \wedge  {o_1 = i_1} \lor \neg c_1 \land {o_1 = neg(i_1)}) \land\\
(c_2\land {o_2=i_2}\lor\neg c_2 \land {o_2 = neg(i_2)})) \limplies
(pos(o_1) \leftrightarrow \neg pos(o_2) ),
\end{multline*}
where the predicate $pos$ returns $\ltrue$ iff its parameter is
positive. We can compute witness functions for $c_1$ and $c_2$ using
$n$-interpolation.\footnote{We must add the axiom
 $(pos(i_1) \oplus pos(neg(i_1))) \land (pos(i_2) \oplus pos(neg(i_2)))$.}
Our method returns witness functions $c_1 = pos(i_1)$ and $c_2 =
\neg pos(i_2)$. (Other functions are also possible.)

Note that computing two interpolants independently may not work. For
instance, we may choose $c_1 = \ltrue$ or we can take $c_2 = \ltrue$,
but we cannot choose $c_1 = c_2 = \ltrue$. This problem is normally
solved by substituting one solution before the next is computed, but
our method computes both interpolants simultaneously and in a
coordinated way.

\section{Preliminaries}
\label{sec:preliminaries}



\subsection{Uninterpreted Functions and Arrays}
We consider the \emph{Theory of Uninterpreted Functions
and Equality} $\theory_U$.
We have variables $\var \in \vars$ from an uninterpreted
domain, Boolean variables $\propvar \in \propvars$, uninterpreted function symbols
$\uf \in \ufs$, and uninterpreted predicate symbols $\up
\in \ups$.
The following grammar defines the syntax of the formulas in
$\theory_u$.
  \begin{align*}
    \text{terms} \ni t & ::=   \var \mid f(t,\dots, t) , & ~\\
    \text{atoms} \ni a & ::=  \propvar \mid P(t,\dots,t) \mid t = t, & ~\\
    \text{formulas} \ni \phi & ::=  a \mid \neg \phi \mid \phi \lor \phi.& ~
  \end{align*}
Let $\phi_1 \land \phi_2$ be short for $\neg (\neg \phi_1 \lor \neg
\phi_2)$. Let $a\neq b$ be short for $\neg(a=b)$. Let $\ltrue=\phi \lor \neg \phi$, let $\lfalse = \neg
\ltrue$, and let $\booleans = \{\ltrue,\lfalse\}$.


A {\em literal} is an atom or its negation. Let $l$ be a literal. If
$l = \neg a$ then let $\neg l = a$.
A {\em clause} is a set of literals, interpreted as the
disjunction. The empty clause $\emptyset$
denotes $\lfalse$.
A {\em conjunctive formula} is the negation of a clause. A {\em CNF
formula} is a set of clauses. A CNF formula is interpreted as the
conjunction of its clauses. Since any formula can be converted into a
CNF formula, we will assume that all the formulas in this paper are
CNF formulas. Let $\phi$ and $\psi$ be CNF formulas/clauses/literals.
Let $Symb(\phi)$ be the set of variables, functions, and predicates
occurring in $\phi$. Let $\phi \preceq \psi $ iff $Symb(\phi)
\subseteq Symb(\psi)$. Let $Lits(\phi) = \{a,\neg a \mid  a \textrm{
is an atom in }\phi \}$. For a clause $C$, let $C|_{\phi} = \{ s \in
C \mid s \preceq \phi\}$.

Arrays are useful for
modeling memory whose size is not known a priori.
We will use a decidable fragment, known as the \emph{Array Property
Fragment} with \emph{uninterpreted
indices}  to create specifications from which we
synthesize controllers.  Bradley~et~al.~\cite{Bradle07} present an
algorithm to reduce formulas with array properties to equisatisfiable
formulas over the theory of uninterpreted functions. Hofferek and
Bloem~\cite{Hoffer11} show that this algorithm generalizes to the
quantified formulas that occur in controller synthesis problems.
For the rest of this paper, we assume that specifications and
formulas containing array properties have been reduced to formulas
over the theory of uninterpreted functions.

\subsection{Proofs of Unsatisfiability}
\begin{figure}[t]
  \footnotesize
%
  \begin{minipage}[t]{0.97\linewidth}
    \centering
    $\inference{\Hyp}{}{\quad C \quad }{C \in \phi,\;\phi \in \text{CNF}}$
    \qquad 
    $\inference{\Axi}{}{\quad C \quad} {\provesTu C} $ \qquad 
    $\inference{\Res}{\quad a \lor C \quad  \neg a \lor D \quad}
    { C \lor D}{}$
    \vspace{1ex}
  \end{minipage}
%
  \vspace{-2ex}
  \caption{
    Sound and complete proof rules for the theory $\theory_U$.
  }
  \vspace{-4ex}
  \label{fig:uf-proof-rules}
\end{figure}


We consider the usual semantics of formulas in $\theory_u$. The
problem of proving unsatisfiability of formulas is decidable.  Many
\emph{Satisfiability Modulo Theories (SMT) Solvers} exist that can
decide the satisfiability of CNF-$\theory_U$ formulas, and, in case
the formula is not satisfiable, produce a \emph{proof of
unsatisfiability}.

A (named) \emph{proof rule} is a template for a logic entailment
between a (possibly empty) list of \emph{premises} and a
\emph{conclusion}.
Templates for premises are written above a horizontal line, templates
for conclusions below. Possible conditions for the application of the
proof rule are written on the right-hand side of the line.

The proofs we consider will be based on the rules given in
Fig.~\ref{fig:uf-proof-rules}. They form a sound and complete proof
system for proving unsatisfiability of a CNF-$\theory_U$ formula
$\phi$. The $\Hyp$ rule is used to introduce clauses from $\phi$ into
the proof. The $\Axi$ rule is used to introduce theory-tautology
clauses. In their simplest form, these clauses represent concrete
instances of theory axioms (reflexivity, symmetry, transitivity and
congruence). However, as our proof transformation algorithms will
produce theory tautologies that are based on several axioms, we use
the following, less restrictive, definition.

\begin{definition}[Theory-Tautology Clause]
A \emph{theory-tautology clause} is a clause of the form $(\neg a_1
\lor \neg a_2 \lor \ldots \lor \neg a_k \lor b)$ that is
tautologically true within the theory $\theory_U$.
The literals $\neg a_i$, for $ 0 < i \leq
k$, are called the \emph{implying literals} and the (positive)
literal $b$ is called the \emph{implied literal}.
\end{definition}

The $\Res$ rule is the standard resolution rule to combine clauses
that contain one literal in opposite polarity respectively. We will
call this literal the \emph{resolving literal} or the \emph{pivot}.

\begin{definition}[Unsatisfiability Proof]
\label{def:proof}
An \emph{unsatisfiability proof} for a CNF-$\theory_U$ formula $\phi$
is a directed, acyclic graph (DAG) $(N, E)$, where $N=\{r\} \cup N_I
\cup N_L$ is the set of nodes (partitioned into the root node $r$,
the set of internal nodes $N_I$, and the set of leaf nodes $N_L$),
and $E\subseteq N \times N$ is the set of (directed) edges. Every $n
\in N$ is labeled with the name of a proof rule $rule(n)$ and a
clause $clause(n)$. The graph has to fulfill the following
properties:
\begin{enumerate}
\item $clause(r)=\emptyset$.
\item For all $n \in N_L$, $clause(n)$ is either a clause from
    $\phi$ (if $rule(n)=\Hyp$) or a theory-tautology clause (if
    $rule(n)=\Axi$).
\item The nodes in $N_L$ whose clauses are theory-tautology
    clauses can be ordered in such a way that for each such node
    each implying literal either occurs in $\phi$, or is an
    implied literals of the tautology clause of a a preceding node (according to the order).\footnote{This means that every
    (new) literal is defined only in terms of previously known
    literals. The order  corresponds to the order in which the solver introduced the
    new literals.}
\item The root has no incoming edges, the leaves have no outgoing
    edges, and all nodes in $n \in N \setminus N_L$ have exactly
    2 outgoing edges, pointing to nodes $n_1, n_2$, with $n_1
    \neq n_2$. Using $clause(n_1)$ and $clause(n_2)$ as premises
    and $clause(n)$ as conclusion must yield a valid instance of
    proof rule $rule(n)$.
\end{enumerate}
\end{definition}
We used the \verit SMT solver \cite{Bouton09}, which provides proofs
that conform to these requirements.

\subsection{Transitivity-Congruence Chains}

Given a set $A$ of atoms, we can use the well-known
congruence-closure algorithm
to construct a \emph{congruence graph} \cite{Fuchs12} according to
the following definition.

\begin{definition}[Congruence Graph]
\label{def:congruence_graph}
A \emph{congruence graph} over a set $A$ of atoms is a graph which
has terms as its nodes. Each edge is labeled either with an
\emph{equality justification}, which is an equality atom from $A$
that equates the terms connected by the edge, or with a
\emph{congruence justification}. A congruence justification can only
be used when the terms connected by the edge are both instances
$\uf(a_1, \ldots, a_k)$ and $\uf(b_1, \ldots, b_k)$ of the same
uninterpreted function $\uf$. In this case, the congruence
justification is a set of $k$ paths in the graph connecting the
$a_i$ and $b_i$ respectively, not using the edge which they label.
\end{definition}

\begin{definition}[Transitivity-Congruence Chain]
\label{def:trans_congr_chain}
A \emph{transitivity-congruence chain} $\pi=(a \leadsto b)$ is a path
in a congruence graph that connects terms $a$ and $b$. Let
$Lits(\pi)$ be the set of literals of the path, which is defined as the
union of the literals of all edges on the path.
The literal of an edge labeled with an equality justification $p$ is
the set $\{p\}$. The set of literals of an edge  labeled with  a
congruence justification with paths $\pi_i$ is recursively defined as
 $\bigcup_i Lits(\pi_i)$. 
\end{definition}

\begin{theorem}
The conjunction of the literals in a transitivity-congruence chain
$(a \leadsto b)$ implies $a=b$ within $\theory_U$. I.e., $(\Lor_{l \in
Lits(a \leadsto b)} \neg l) \lor (a=b)$ is a theory-tautology clause.
\end{theorem}

\subsection{Craig Interpolation}
\begin{figure}[t]
  \footnotesize
  \begin{minipage}[t]{1.0\linewidth}
    \centering
  $\inference{\iHypX}{}{
    \quad C[\lfalse] \quad} {C \in \phi }$ \qquad
  $\inference{\iHypY}{}{\quad C[\ltrue] \quad} {C \in \psi }$ \\
  \vspace{1ex}
  $\inference{\iAxiX}{}
  {\quad C[\lfalse] \quad} { C \preceq \phi, \provesTu C }$
  \qquad
  $\inference{\iAxiY}{}
  {\quad C[\ltrue] \quad} { C \preceq \psi, \provesTu C }$
  \\
  \vspace{1ex}
  $\inference{\iResXY}
  { a \lor C[I_C] \qquad
     \neg a \lor D[I_D] }
  { \quad C \lor D[(a \lor I_C)\land (\neg a \lor I_D)] \quad}
  {a \preceq \phi, a \preceq \psi}$\\
  \vspace{1ex}
  $\inference{\iResX}{\quad a \lor C[I_C] \qquad \neg a \lor D[ID] \quad}
  {\quad C \lor D[I_C \lor I_D] \quad}
  {a \preceq \phi, a \npreceq \psi}$\\
  \vspace{1ex}
  $\inference{\iResY}{
     \quad a \lor C[I_C] \qquad
    \neg a \lor D[I_D] \quad}
  {\quad C \lor D[I_C \land I_D] \quad}
  {a \npreceq \phi, a \preceq \psi}$ \\
  \end{minipage}
  \vspace{-3ex}
  \caption{
    Interpolating proof rules
  }
  \vspace{-3ex}
  \label{fig:interpolation}
\end{figure}


Let $\phi$ and $\psi$ be CNF formulas such that $\phi \land \psi$ is
unsatisfiable. The algorithm presented in~\cite{McMillanSplitProver}
for computing an interpolant between $\phi$ and $\psi$ needs a proof
of unsatisfiability of $\phi \land \psi$. By annotating this proof
with the partial interpolants, the algorithm computes the
interpolant. In this paper, we present slightly different annotation
rules to compute interpolants, which are results of mixing ideas
from~\cite{KovacsCADE09,Pudlak97}.

\begin{definition}[Partial interpolant]
Let $C$ be a clause such that $\phi \land \psi \limplies C$. A
formula $I$ is a {\em partial interpolant} for $C$ between $\phi$ and
$\psi$ if $\phi \limplies C|_{\phi} \lor I$, and $\psi \limplies
C|_{\psi} \lor \neg I$, and $I \preceq \phi, \text{ and } I \preceq
\psi$.
If $I$ is a partial interpolant for $C = \emptyset$ between $\phi$ and $\psi$,
then $I$ is an interpolant between $\phi$ and $\psi$.
\end{definition}

In Figure~\ref{fig:interpolation}, we present interpolating proof
rules. In an unsatisfiability proof of $\phi \land \psi$, these rules
annotate (in square brackets) each conclusion with a partial
interpolant for the conclusion. Rules $\iHypX$ and $\iHypY$ are used
at leaf nodes that have clauses from $\phi$ and $\psi$ respectively.
Rules $\iAxiX$ and $\iAxiY$ are used for leaves with theory-tautology
clauses, whose symbols are a subset of the symbols in $\phi$ and
$\psi$ respectively. Note that these rules assume that the
unsatisfiability proof of $\phi \land \psi$ is {\em colorable}.

\begin{definition}[Colorable Proof]
\label{def:localized-proof}
A proof of unsatisfiability of $\phi \land \psi$ is \emph{colorable}
if for every leaf $n_L$ of the proof $Symb(clause(n_L)) \subseteq
Symb(\phi)$ or $Symb(clause(n_L)) \subseteq Symb(\psi)$.
\end{definition}
In Section~\ref{sec:localize}, we will present an algorithm that
transforms a proof into a colorable proof. Due to this assumption we
can easily find corresponding partial interpolants for
theory-tautology clauses, which are either $\ltrue$ or $\lfalse$. For
internal proof nodes, we follow Pudl\'ak's interpolation
system~\cite{Pudlak97}. The annotation of the root node (with the
empty clause) is the interpolant between $\phi$ and $\psi$.
See~\cite{DSilvaStengthVMCAI10} for a proof of correctness of the
annotating proof rules.


\section{Controller Synthesis}
\label{sec:synthesis}

\subsection{Overview}
\label{sec:synthesis-overview}

Following \cite{Hoffer11}, we assume that synthesis problems are
given as formulas of the form
\begin{equation}
  \label{eq:synth}
  \forall \is \; \exists \cs\; \forall \os. \; \phiValid(\is,\cs,\os),
\end{equation}
where $\cs$ is a vector of Boolean variables and $\phiValid$ is a
formula over theory~$\theory_U$. Let $\cs = (c_1,\dots,c_n)$. Each
$c_i$ represents a missing if-condition in a program or a one-bit
control signal in a hardware design.
Witness functions for the existentially quantified variables in
Eq.~\eqref{eq:synth} are implementations of the missing components.
Therefore, the synthesis problem is equivalent to finding such
witness functions. I.e., find $(f_1(\is),\dots,f_n(\is))$ such that
$\forall \is\; \forall \os. \;
\phiValid(\is,(f_1(\is),\dots,f_n(\is)),\os)$ holds true.

We compute the witness functions through the following steps:
\begin{enumerate}
\item Expand the existential quantifier and negate the formula
    $\phiValid$ to obtain an unsatisfiable formula $\phiUnsat$
    (Sec.~\ref{sec:interpol-synth}).
\item Obtain a proof of unsatisfiability from an SMT solver.
\item Transform the proof into a colorable, \lfirst~proof
    (Sec.~\ref{sec:localize}).
\item Perform $n$-interpolation on the transformed proof. The
    elements of the $n$-interpolant correspond to the witness
    functions (Sec.~\ref{sec:interpol-synth}).
\end{enumerate}

We will first introduce the notion of $n$-interpolation and show how
it is used to find witness functions in
Section~\ref{sec:interpol-synth}. Subsequently, we will show how to
transform a proof of unsatisfiability so that it is suitable for
$n$-interpolation in Section~\ref{sec:localize}.

\subsection{Finding Witness Functions through Interpolation}
\label{sec:interpol-synth}

Jiang~et~al.~\cite{JiangICADD09} show how to compute a witness
function in Eq.~\eqref{eq:synth} using interpolation if $\cs$
contains a single Boolean $c$. In this case, Eq.~\eqref{eq:synth}
reduces to $\forall \is \; \exists c\; \forall \os. \;
\phiValid(\is,c,\os).$ After expanding the existential quantifier by
instantiating the above formula for both Boolean values of $c$ and
renaming $\os$ in each instantiation, we obtain the equivalent
formula $ \forall \is \; \forall \os_\lfalse,\os_\ltrue. \;
\phiValid(\is,\lfalse,\os_\lfalse) \lor
\phiValid(\is,\ltrue,\os_\ltrue)$. Since all the quantifiers are
universal, the disjunction is valid. Therefore, its negation
$\neg\phiValid(\is,\lfalse,\os_\lfalse) \land \neg
\phiValid(\is,\ltrue,\os_\ltrue)$ is unsatisfiable. The interpolant
between the two conjuncts is the witness function for variable~$c$.
\begin{theorem}
 \label{thm:intpol-is-witness}
The interpolant between $\neg\phiValid(\is,\lfalse,\os_\lfalse)$ and
$\neg \phiValid(\is,\ltrue,\os_\ltrue)$ is the witness function for
$c$. (For a proof, see
\iftoggle{cameraready}{\cite{fullversion}}{Appendix~\ref{proof:intpol-is-witness}}.)
\end{theorem}

We now extend this idea to compute witness functions when $\cs$ is a
vector of Booleans $(c_1,\dots,c_n)$.
%
Let $\booleans^n$ denote the set of vectors of length $n$ containing
$\lfalse$s and $\ltrue$s. For vector $w \in \booleans^n$, let $w_j$
be the Boolean value in $w$ at index~$j$. Since $\cs$ is a Boolean
vector, we can expand the existential quantifier for $\cs$ in
Eq.~\eqref{eq:synth} by enumerating the finitely many possible values
of $\cs$ to obtain
$\forall \is \Lor_{w \in \booleans^n}\forall \os\; .
\;\phiValid(\is,w ,\os)$. By dropping the quantifiers and renaming
$\os$ accordingly, we obtain
$\Lor_{w \in \booleans^n} \phiValid(\is,w,\os_w)$.
It is valid iff Eq.~\eqref{eq:synth} is valid. Let $\phiUnsat$ denote
its negation $ \Land_{w \in \booleans^n} \neg
\phiValid(\is,w,\os_w)$, which is unsatisfiable. Let $\phiUnsat_w$
denote $\neg \phiValid(\is,w,\os_w)$.
We will call the $\phiUnsat_w$s the $2^n$ \emph{partitions} of
$\phiUnsat$. We will learn a vector of {\em coordinated} interpolants
from an unsatisfiability proof of $\phiUnsat$. These interpolant
formulas will be witness functions for $\cs$. Since $\phiUnsat_w$s
are obtained by only renaming variables, the shared symbols between
any two partitions are equal.

\begin{definition}[Global and Local Symbols]
Symbols in the set $\commonLits = \bigcap_{w \in \booleans^n}
Symb(\phiUnsat_w)$ are called \emph{global} symbols. All other
symbols are called \emph{local} (w.r.t. the one partition in which
they occur).
\end{definition}

Let $\vecI$ be a vector of formulas $(I_1,\dots,I_n)$. Let $\xor$ be
the exclusive-or (xor) operator. For a word $w \in \booleans^n$, let
$\vecI' = \vecI \xor w$ if for each $j \in \betweenOf{1}{n}$, $I'_j =
I_j \xor w_j $. Let $\Lor \vecI$ be short for $I_1 \lor \dots \lor
I_n$. Let $C|_w = C|_{\phiUnsat_w}$. The following definition
generalizes the notion of interpolant and partial interpolant from
two formulas to $2^n$ formulas.

\begin{definition}[$n$-Partial Interpolant]
\label{def:n-part-interpol} Let $C$ be a clause such that $(\Land_{w
\in \booleans^n} \phiUnsat_w) \limplies C$. An n-partial
interpolant~$\vecI$ for $C$ w.r.t. the~$\phiUnsat_w$s is a vector of
formulas with length $n$ such that $ \forall w \in \booleans^n.\;
\phiUnsat_w \limplies (C|_w \lor \Lor (\vecI\xor w))$ and $\vecI
\preceq G.$ If $C = \emptyset$ then $\vecI$ is an {\em n-interpolant}
w.r.t.~the~$\phiUnsat_w$s.
\end{definition}

\begin{theorem}
  \label{thm:n-intpol-are-witnesses}
  An $n$-interpolant w.r.t.~the~$\phiUnsat_w$s constitutes
  witness functions for the variables in $\cs$.
(For a proof see
\iftoggle{cameraready}{\cite{fullversion}}{Appendix~\ref{proof:n-intpol-are-witnesses}}.)
\end{theorem}

\begin{figure}[t]
  \footnotesize
  \begin{minipage}[t]{1.0\linewidth}
    \centering
  $\inference{\mHyp}{}{
    \quad C[w] \quad} {C \in \phiUnsat_w }$ \quad
  $\inference{\mAxi}{}
  {\quad C[w] \quad} { C \preceq \phiUnsat_w }$ \\
  \vspace{1ex}
  $\inference{\mRes}{\quad a \lor C[w]
    \qquad \neg a \lor D[w] \quad}
  { C \lor D[w]}
  {
   \begin{array}{@{}l@{}}
     w \in \booleans^n ,
      a \lor C \lor D\preceq \phiUnsat_w
   \end{array}
  }$\\
  \vspace{1ex}
  $\inference{\mResG}
  {\quad a \lor C[\vecI^C] \qquad
     \neg a \lor D[\vecI^D] \quad}
  {C \lor D
     \begin{array}[t]{c@{\;}c@{\;}l}
       [( &(a \lor I_1^C)\land (\neg a \lor I_1^D),& \\
          &\dots,&\\
          &(a \lor I_n^C)\land (\neg a \lor I_n^D)& )]
     \end{array}
  }
  {a \preceq G}$\\
  \end{minipage}
  \vspace{-2ex}
  \caption{
    $n$-Interpolating proof rules for an unsatisfiable
    $\phi = \Land_{w \in \booleans^n} \phiUnsat_w$.
    These rules can only annotate proofs that
    are colorable and \lfirst.
  }\vspace{-4ex}
  \label{fig:multi-interpolation}
\end{figure}


\subsection{Computing $n$-interpolants}
In Figure~\ref{fig:multi-interpolation}, we present the proof rules
for $n$-interpolants. 
These rules annotate each conclusion of a proof step with an
$n$-partial interpolant for the conclusion w.r.t.~the~$\phiUnsat_w$s.
These annotation rules require two properties of the proof. First, it
needs to be colorable.\footnote{We extend
Def.~\ref{def:localized-proof} from two formulas to $2^n$ partitions
in the obvious way.} Second, it needs to be \lfirst.
\begin{definition}[Local-first Proof]
A proof of unsatisfiability is \emph{\lfirst}, if for every
resolution node with a local pivot both its premises are derived from
the same partition.
\end{definition}

The rule $\mHyp$ annotates the derived clause $C$ with $w$ if $C$
appears in partition $\phiUnsat_w$. Similarly, the rule $\mAxi$
annotates theory-tautology clause $C$ with $w$ if $C \preceq
\phiUnsat_w$. Rules $\mRes$ and $\mResG$ annotate resolution steps.
$\mResG$ is only applicable if the pivot is global and follows
Pudl\'ak's interpolation system $n$ times. $\mRes$ is only applicable
if both premises are annotated with the same $n$-partial interpolant
and this $n$-partial interpolant is an element of $\booleans^n$. Due
to the \lfirst~assumption on proofs, these rules will always be able
to annotate a proof.

\begin{theorem}
  \label{thm:annotations-are-interpolants}
  Annotations in the rules in Figure~\ref{fig:multi-interpolation} are
  $n$-partial interpolants for the respective conclusions w.r.t.~the~$\phiUnsat_w$s.
  (For a proof see \iftoggle{cameraready}{\cite{fullversion}}{Appendix~\ref{proof:annotations-are-interpolants}}.)
\end{theorem}

Since the $n$-interpolant is always quantifier free, we can easily
convert it into an implementation. To create a circuit for one
element of the $n$-interpolant, we create, for every resolution node
with a global pivot, a multiplexer that has the pivot at its selector
input. The other inputs connect to the outputs of the multiplexers
corresponding to the child nodes. For leaf nodes and resolution nodes
with local pivots, we use the constants $\ltrue, \lfalse$, depending
on which partition the node belongs to. The output of the multiplexer
corresponding to the root node is the final witness function. Note
that, unless we apply logical simplifications, the circuits for all
witness functions all have the same multiplexer tree and differ only
in the constants at the leaves of this tree.

Also note that due to the \lfirst~property, all nodes that are
derived from a single partition are annotated with the same
$n$-partial interpolant. Thus, we can disregard such local sub-trees,
by iteratively converting nodes that have only descendants from one
partition into leaves. This does not affect the outcome of the
interpolation procedure.

The \lfirst~property is actually needed to correctly compute witness
functions using Pudl\'ak's interpolation system. In
\iftoggle{cameraready}{\cite{fullversion}}{Appendix~\ref{why-local-first}},
we illustrate this observation with an example. Also note that
McMillan's interpolation~\cite{McMillanSplitProver} system does not
produce correct witness functions even with the \lfirst~property.


\section{Algorithms for Proof Transformation}
\label{sec:localize}

Our interpolation procedure requires proofs to be colorable and
\lfirst. These properties are not guaranteed by efficient modern SMT
solvers. In this section we will show how to transform a proof
conforming to Def.~\ref{def:proof} into one that is colorable and
\lfirst.
Our proof transformation works in three steps. First, we will remove
any non-colorable literals from the proof. Second, we will split any
non-colorable theory-tautology clauses. This gives us a colorable
proof. In the third step, we will reorder resolution steps to obtain
the \lfirst~property \cite{DSilvaStengthVMCAI10}. For ease of
presentation, we will assume that the proof is a tree (instead of a
DAG). The method extends to proofs in DAG form.

\subsection{Removing Non-Colorable Literals}

\begin{figure*}[t]

  \centering
  \scriptsize
\addtolength{\subfigcapskip}{0.3cm}
\subfigure[Proof before removing non-colorable literal $l_1=l_2$.]{
 $
    \inference{\mathllap\Res}{
     \inference{\Res}{ n_1:( l_1=z_g \lor  x_g=y_g)  \;\;   n_d: ( l_1\neq z_g \lor z_g\neq l_2 \lor l_1=l_2) }
    {n_a:(x_g=y_g \lor z_g\neq l_2 \lor l_1=l_2 )}{}
    \;
    \inference{\Res}{ n_u:(l_1\neq l_2 \lor f(l_1)=f(l_2)) \;\; n_3:(f(l_1)\neq f(l_2) \lor u_g \neq v_g)  }
    {n_{\neg a}:(l_1\neq l_2 \lor u_g \neq v_g)}{}
    }{n_r:(x_g=y_g \lor z_g\neq l_2 \lor u_g \neq v_g)}{}
$
   \label{fig:remove_non_local_before}
}

\subfigure[Proof after removing non-colorable literal $l_1=l_2$.]{
  $
    \inference{\mathllap\Res}{
   n_1: ( l_1=z_g \lor  x_g=y_g)
    \;
    \inference{\Res}{ n_3:(f(l_1)\neq f(l_2) \lor u_g \neq v_g) \;\; n'_u:(l_1 \neq z_g \lor z_g \neq l_2 \lor f(l_1) = f(l_2))  }
    {n'_{\neg a}:(l_1\neq z_g \lor z_g \neq l_2 \lor u_g \neq v_g)}{}
    }{n''_{\neg a}:(x_g=y_g \lor z_g\neq l_2 \lor u_g \neq v_g)}{}
  $

   \label{fig:remove_non_local_after}
}

\caption{\emph{Removing a non-colorable literal.} Assume that term
indices indicate the number of the partition the term belongs to.
Index $g$ is used for global terms. This example shows how the
non-colorable literal $l_1 = l_2$, introduced in node $n_d$, is
removed from the proof by replacing its negative occurrences with the
(colorable) defining literals $(l_1\neq z_g \vee z_g \neq l_2)$. Note
that in the original proof $l_1\neq z_g$ is already resolved on the
path from $n_d$ to $n_r$ using node $n_1$. This resolution step is
replicated in the transformed proof by making a resolution step with
nodes $n'_{\neg a}$ and $n_1$. Since the literal $x_g=y_g$ introduced
into $n''_{\neg a}$ also occurs in the original $n_r$, and also the
second defining literal $z_g \neq l_2$ occurs in $n_r$, no further
resolution steps are necessary. The conclusions of $n_d$ and
$n''_{\neg a}$ are identical and $n''_{\neg a}$ can be used instead
of $n_r$ in $n_r$'s parent.}
\label{fig:remove_non_local} %
\vspace{-5ex}
\end{figure*}


\begin{definition}[Colorable and Non-Colorable Literals]
\label{def:non-local-literal} %
A literal $a$ is \emph{colorable}  with
respect to a partition $\phiUnsat_w$ (\emph{$w$-colorable}) iff $a \preceq \phiUnsat_w$. A
literal that is not $w$-colorable for any partition $w$ is called
\emph{non-colorable}.
\end{definition}
Note that global literals are
$w$-colorable for every $w$. By definition, the formula $\phiUnsat$ is free of non-colorable
literals (equalities and predicate instances).
Thus, the only way through which non-colorable literals can be
introduced into the proof are theory-tautology clauses.

We search the proof for a theory-tautology clause that introduces a
non-colorable literal $a$ and has only colorable literals as implying
literals. We call this proof node the \emph{defining node} $n_d$. At
least one such leaf must exist. We remove this non-colorable literal
from the proof as follows.
Starting from $n_d$, we traverse the proof towards the root, until we
find a node, which we call \emph{resolving node} $n_r$, whose clause
does not contain the literal $a$ any more. Since the root node does
not contain any literals, such a node always exists. Let $n_a$ and
$n_{\neg a}$ be the premises of $n_r$, respectively, depending on
which phase of literal $a$ their clause contains.
From $n_{\neg a}$, we traverse the proof towards the leaves along
nodes that contain the literal $\neg a$. Note that any leaf that we
reach in this way must be labeled with a theory-tautology clause, as
clauses from $\phiUnsat$ cannot contain the non-colorable literal
$\neg a$. Note that $\neg a$ is among the implying literals of such a
leaf node's clause. I.e., such nodes \emph{use} the literal to imply
another one. We will therefore call such a node a \emph{using node}
$n_u$. We update $clause(n_u)$, by removing $\neg a$ and adding the
implying literals of $clause(n_d)$ instead.

It is easy to see that this does not affect $clause(n_u)$'s property
of being a theory-tautology clause. Suppose $clause(n_d)$ is $(\neg
x_1 \vee \ldots \vee \neg x_k \vee a)$. Then $\bigwedge_{i=1}^k x_i
\rightarrow a$. By reversing the implication we obtain $\neg a
\rightarrow \bigvee_{i=1}^k \neg x_i$. Therefore, replacing $\neg a$
with the disjunction of the implying literals of $clause(n_d)$ in
$clause(n_u)$ is sound.

To keep the proof internally consistent, we have to do the same
replacement on all the nodes on the path between $n_u$ and $n_r$. The
node $n_r$ itself is not changed, as $clause(n_r)$ does not contain
the non-colorable literal $(\neg)a$ any more. I.e., the last node
that is updated is the node $n_{\neg a}$.

Now we have to distinguish two cases. The first case is that node
$n_a$ still contains all of the implying literals of $n_d$. In this
case, $clause(n_r)=clause(n'_{\neg a})$, where $n'_{\neg a}$ is the
updated node $n_{\neg a}$. Thus, we use $n'_{\neg a}$ instead of
$n_r$ in $n_r$'s parent node.
In the second case, some of the implying literals of $clause(n_d)$
have already been resolved on the path from $n_d$ to $n_r$. In that
case $clause(n'_{\neg a})$ contains literals that do not occur in
$clause(n_r)$. Let $x_l$ be one such literal. We search the path from
$n_d$ to $n_r$ for the node that uses $x_l$ as a pivot. Its premise
that is not on the path from $n_d$ to $n_r$ contains $\neg x_l$. We
use this node and the node $n'_{\neg a}$ as premises for a new
resolution node with $x_l$ as pivot. Note that this resolution may
introduce more literals that do not appear in $clause(n_r)$ any more.
However, just as with $x_l$, any such literal must have been resolved
somewhere on the path between $n_d$ and $n_r$. Thus, we  repeat this
procedure, replicating the resolution steps that took place between
$n_d$ and $n_r$, until we get a node whose clause is identical to
$clause(n_r)$. This node can then be used instead of $n_r$ in $n_r$'s
parent node. Finally, we remove all nodes that are now unreachable
from the proof.

\begin{example}
An illustrative example of this procedure is shown in
Figure~\ref{fig:remove_non_local}.
\end{example}

We repeat this procedure for all leaves with a non-colorable implied
literal and (all) colorable implying literals. Note that one
application of this procedure may convert a node where a
non-colorable literal was implied by at least one other non-colorable
literal into a node where the implied non-colorable literal is now
implied only by colorable literals. Nevertheless this procedure
terminates, as the number of leaves with non-colorable implied
literals decreases with every iteration. Each iteration removes (at
least) one such leaf from the proof and no new leaves are introduced.

\begin{theorem}
Upon termination of this procedure, the proof does not contain any
non-colorable literals.
\end{theorem}

\subsection{Splitting Non-Colorable Theory-Tautology Clauses}

After removing all non-colorable literals, the proof may still
contain non-colorable theory-tautology clauses, 
i.e., theory-tautology clauses that contain local literals from more
than one partition. We split such leaves into several new
theory-tautology clauses, each containing only $w$-colorable
literals, and, via resolution, obtain a (now internal) node with the
same clause as the original non-colorable theory-tautology clause.
Note that internal nodes with non-colorable clauses are not a problem
for our interpolation procedure, but leaves with non-colorable
clauses are.
We will show how to split a non-colorable theory-tautology clause
with an implied equality literal. This procedure can be trivially
extended to implied literals that are uninterpreted predicate
instances.

Using the implying literals of the theory-tautology clause (converted
to their positive phase), we create a \emph{congruence graph}
(cf.~Def.~\ref{def:congruence_graph}). Since the implying literals
and the implied literal form a theory tautology, this congruence
graph is guaranteed to contain a path between the the two terms
equated by the implied literal. We use breadth-first search to find
the shortest such transitivity-congruence chain
(Def.~\ref{def:trans_congr_chain}).\footnote{Note that these graphs
are usually relatively small.} The chain will be the basis for
splitting the non-colorable theory tautology. First, we need to make
all edges in the chain colorable. A \emph{colorable edge} is an edge
for which there is a $w$ such that all the edge's literals are
$w$-colorable. Edges with an equality justification already are
colorable, as we assumed that no non-colorable literals occur in the
theory-tautology clause. Edges with congruence justifications,
however, may still be non-colorable. I.e., the two terms they connect
might belong to different partitions, and/or some of the paths that
prove equality for the function parameters might span over more than
one partition. Fuchs~et~al.~\cite{Fuchs12} have shown how to
recursively make all edges in a chain colorable by introducing global
intermediate terms for non-colorable edges. We will illustrate this
procedure with a simple example, and refer to \cite{Fuchs12} for
details.

\begin{figure}[t]
\centering

\subfigure[Non-Colorable Transitivity-Congruence Chain for $(f(l_1)
\leadsto f(l_2))$]{
   \includegraphics[scale=0.55] {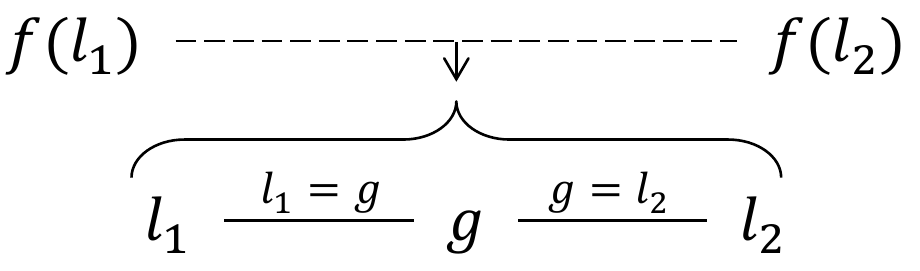}
   \label{fig:subfig1}
 }

 \subfigure[Colorable Transitivity-Congruence Chain for $(f(l_1)
\leadsto f(l_2))$]{
   \includegraphics[scale=0.55] {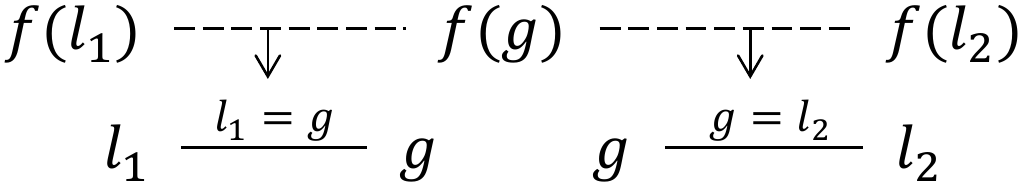}
   \label{fig:subfig2}
 }
\caption{Splitting a non-colorable transitivity-congruence chain by
introducing global intermediate terms.} \label{fig:splitting_chain}
\end{figure}
\begin{figure*}[t]

  \centering
  \scriptsize

  $$
    \inference{\Res}{
     \inference{\Res}{ n_1:[c_g \neq d_2 \lor d_2 \neq e_2 \lor e_2 \neq f_g \lor f_g \neq k_g \lor c_g = k_g]  \;\;   n_2: [ f_g \neq h_3 \lor h_3 \neq k_g \lor f_g = k_g] }
    {n_3:[c_g \neq d_2 \lor d_2 \neq e_2 \lor e_2 \neq f_g \lor f_g \neq h_3 \lor h_3 \neq k_g \lor c_g=k_g ]}{}
    \;
  n_4:[a_1\neq b_1 \lor b_1 \neq c_g \lor c_g \neq k_g \lor k_g \neq l_1 \lor a_1 = l_1]
    }{n_5:[a_1 \neq b_1 \lor b_1 \neq c_g \lor c_g \neq d_2 \lor d_2 \neq e_2 \lor e_2 \neq f_g \lor f_g \neq h_3 \lor h_3 \neq k_g \lor k_g \neq l_1 \lor a_1=l_1]}{}
  $$

\caption{ %
\emph{Splitting theory tautology clauses.} Suppose we have created
the transitivity-congruence chain $(a_1 \leadsto b_1 \leadsto c_g
\leadsto d_2 \leadsto e_2 \leadsto f_g \leadsto h_3 \leadsto k_g
\leadsto l_1)$ from a theory-tautology clause, where all the edges
are colorable. The number in the index indicates the partition of the
respective term, with $g$ being used for global terms. First, we
consider only the part from the first to the last global term ($c_g$
and $k_g$, respectively). We ``split'' this sub-chain into the chains
$(c_g \leadsto d_2 \leadsto e_2 \leadsto f_g \leadsto k_g)$ and $(f_g
\leadsto h_3 \leadsto k_g)$ and convert them into (colorable) theory
tautology clauses (nodes $n_1$ and $n_2$, respectively). By
resolution we obtain $n_3$. Now, we create the tautology in node
$n_4$, which corresponds to all links of the original chain which we
have not dealt with already, and a ``shortcut'' over the part we have
already considered: $(a_1 \leadsto b_1 \leadsto c_g \leadsto k_g
\leadsto l_1)$. Note that this is also a colorable theory-tautology
clause. By resolution over $n_3$ and $n_4$ we obtain $n_5$, whose
clause is identical to the theory-tautology clause from which we
started.
 }
\label{fig:splitting_clauses} %
\vspace{-5ex}
\end{figure*}


\begin{example}
Suppose we have the two local terms $f(l_1)$ and $f(l_2)$, where
$l_1, l_2$ are from two different partitions, and a global term $g$.
(See Fig.~\ref{fig:splitting_chain}.) A possible (non-colorable)
congruence justification for $f(l_1)=f(l_2)$ could be given as
$(l_1=g,g=l_2)$. The edge between $f(l_1)$ and $f(l_2)$ is now split
into two (colorable) parts: $f(l_1)=f(g)$, with justification
$l_1=g$, and $f(g)=f(l_2)$, with justification $g=l_2$. Note that
$f(g)$ is a new term that (possibly) did not appear in the congruence
graph before. Since we assumed that there are no non-colorable
equality justifications in our graph, such a global intermediate term
must always exist. It should be clear how to extend this procedure to
$n$-ary functions.
\end{example}

Note that in a colorable chain, every edge either connects two terms
of the same partition, or a global term and a local term. In other
words, terms from different partitions are separated by at least one
global term between them. We now divide the whole chain into
(overlapping) segments, so that each segment uses only $w$-colorable
symbols. The global terms that separate symbols with different colors
are part of both segments.\footnote{If there is more than one
consecutive global term, we arbitrarily choose the last one.}
Let's assume for the moment that the chain starts and ends with a
global term. We will show how to deal with local terms at the
beginning/end of the chain later. For ease of presentation, also
assume that the chain consists of only two segments. An extension to
chains with more segments can be done by recursion. We take the first
segment of the chain (from start to the global term that is at the
border to the next segment), plus a new ``shortcut'' literal that
states equality between the last term of the first segment and the
last term of the entire chain, and use them as implying literals for
a new theory-tautology clause. The implied literal of this tautology
will be an equality between the first and the last term of the entire
chain. Next, we create a theory tautology with the literals of the
second segment of the chain. Note that the implied literal of this
theory-tautology clause (which occurs in positive phase) is the same
as the shortcut literal used in the theory-tautology clause
corresponding to the first segment. There, however, it occurred in
negative phase. Thus, we can use this literal for resolution between
the two clauses. We obtain a node that has all the literals of the
entire chain as implying literals, and an equality between start term
and end term of the chain as the implied literal. I.e., this new
internal node has the same conclusion as the non-colorable
theory-tautology clause from which we started.

In case the start/end of the chain is not a local term, we first deal
with the sub-chain from the first to the last global term, as
described above. Note that if both start and end of the chain are
local terms, they have to belong to the same partition, because
otherwise the implied literal would be non-colorable.
We create a theory-tautology clause with the local literals
from the start/end of the chain, and one shortcut literal that
equates the first and last global term. This literal can be used for
resolution with the implied literal of the node obtained in the
previous step.

In summary, this procedure replaces all leaves that have
non-colorable theory-tautology clauses with subtrees whose leaves are
all colorable theory-tautology clauses, and whose root is labeled
with the same clause as the original non-colorable leaf.

\begin{example}
Fig.~\ref{fig:splitting_clauses} shows how to split the non-colorable
theory-tautology clause $({a_1 \neq b_1} \lor {b_1 \neq c_g} \lor
{c_g \neq d_2} \lor {d_2 \neq e_2} \lor {e_2 \neq f_g} \lor {f_g \neq
h_3} \lor {h_3 \neq k_g} \lor {k_g \neq l_1} \lor {a_1=l_1})$.
\end{example}

\begin{theorem}
After applying the above procedure to all leaves with non-colorable
theory-tautology clauses, the proof is colorable.
\end{theorem}

\subsection{Obtaining a \lfirst~proof}
\begin{figure*}[t]
  \centering
  \scriptsize
  $$
  \begin{array}{@{}l@{}}
  \begin{array}{@{}l@{}}
    \inference{\Res}{ g \lor l \lor D \quad \neg g \lor E }{}{}
    \vspace{-2.5ex}\\
    \hspace{5ex}\inference{\Res}{
        l \lor D\lor E \quad \neg l \lor C
    }{C \lor D \lor E}{}
  \end{array}
  \begin{array}{@{\;}c@{\;}}\mbox{}\\\rightsquigarrow\end{array}
  \begin{array}{@{}l@{}}
    \inference{\Res}{ g \lor l \lor D \quad  \neg l \lor C}{}{}
    \vspace{-2.5ex}\\
    \hspace{5ex}\inference{\Res}{g \lor C\lor D \quad \neg g \lor E
    }{C \lor D \lor E}{}
  \end{array}
  \\
  \begin{array}{@{}l@{}}
    \inference{\Res}{ g \lor l \lor D \quad \neg g \lor l\lor E }{}{}
    \vspace{-2.5ex}\\
    \hspace{6ex}\inference{\Res}{
      l \lor D\lor E \quad \neg l \lor C
    }{C \lor D \lor E}{}
  \end{array}
  \begin{array}{@{\;}c@{\;}}\mbox{}\\\rightsquigarrow\end{array}
  \begin{array}{@{}l@{}}
    \inference{\Res}{ g \lor l \lor D  \;\; \neg l \lor C }
    {g \lor C \lor D}{}
    \;
    \inference{\Res}{ \neg g \lor l\lor E \;\;  \neg l \lor C  }
    {\neg g \lor C \lor E}{}
    \vspace{-2.5ex}\\
    \hspace{8ex}\inference{\mathllap\Res}{\hspace{38ex}}{C \lor D \lor E}{}
  \end{array}
  \end{array}
  $$
\vspace{-6ex}
  \caption{If a local pivot $l$ occurs after a global pivot $g$ in a proof then
    we can rewrite the proof using one of the above
    transformation rules.
    After the transformation, the proof first resolves $l$ then $g$.
  }
\vspace{-6ex}
  \label{fig:transform-res}
\end{figure*}


To obtain a \lfirst~proof, we traverse the proof in topological
order. Each time we encounter a resolution step that has a global
pivot and we have seen local pivots among its ancestors then we apply
one of the two transformation rules presented in
Figure~\ref{fig:transform-res} depending on the matching pattern.
These two transformation rules are the standard pivot reordering
rules from~\cite{DSilvaStengthVMCAI10}. Note that these rules assume
that the proof is redundancy free, which can be achieved by the
algorithms presented in~\cite{GuptaATVA12}. After repeated
application of these transformation rules, we can move the
resolutions with local pivots towards the leaves of the proof until
we don't have any global pivot among its descendants.
\begin{theorem}
After exhaustive application of this transformation, we obtain a
colorable, \lfirst~proof.
\end{theorem}


\section{Experimental Results}
\label{sec:experiments}

We have implemented a prototype to evaluate our interpolation-based
synthesis approach. We read the formula $\phiValid$ corresponding to
our synthesis problem (Eq.~\ref{eq:synth}) from a file in SMT-LIB
format~\cite{Barret10}. As a first step, our tool performs several
transformations on the input formula (reduction of arrays to
uninterpreted functions~\cite{Bradle07}, expansion of the existential
quantifier to obtain the partitions, renaming of $\os$-variables in
each partition, negation to obtain $\phiUnsat$), before giving it to
the \verit solver. Second, we apply the proof transformations
described in Section~\ref{sec:localize} to the proof we obtain from
\verit. Third, we compute the witness functions as the
$n$-interpolants w.r.t.~the partitions of $\phiUnsat$.
%
%

We have checked all results using \Zthree~\cite{Z3}, by showing that
$\neg \phiValid(\is,(f_1(\is),\dots,f_n(\is)),\os)$ is unsatisfiable.

\begin{figure}[t]
  \begin{center}
    \includegraphics[width=0.4\textwidth]{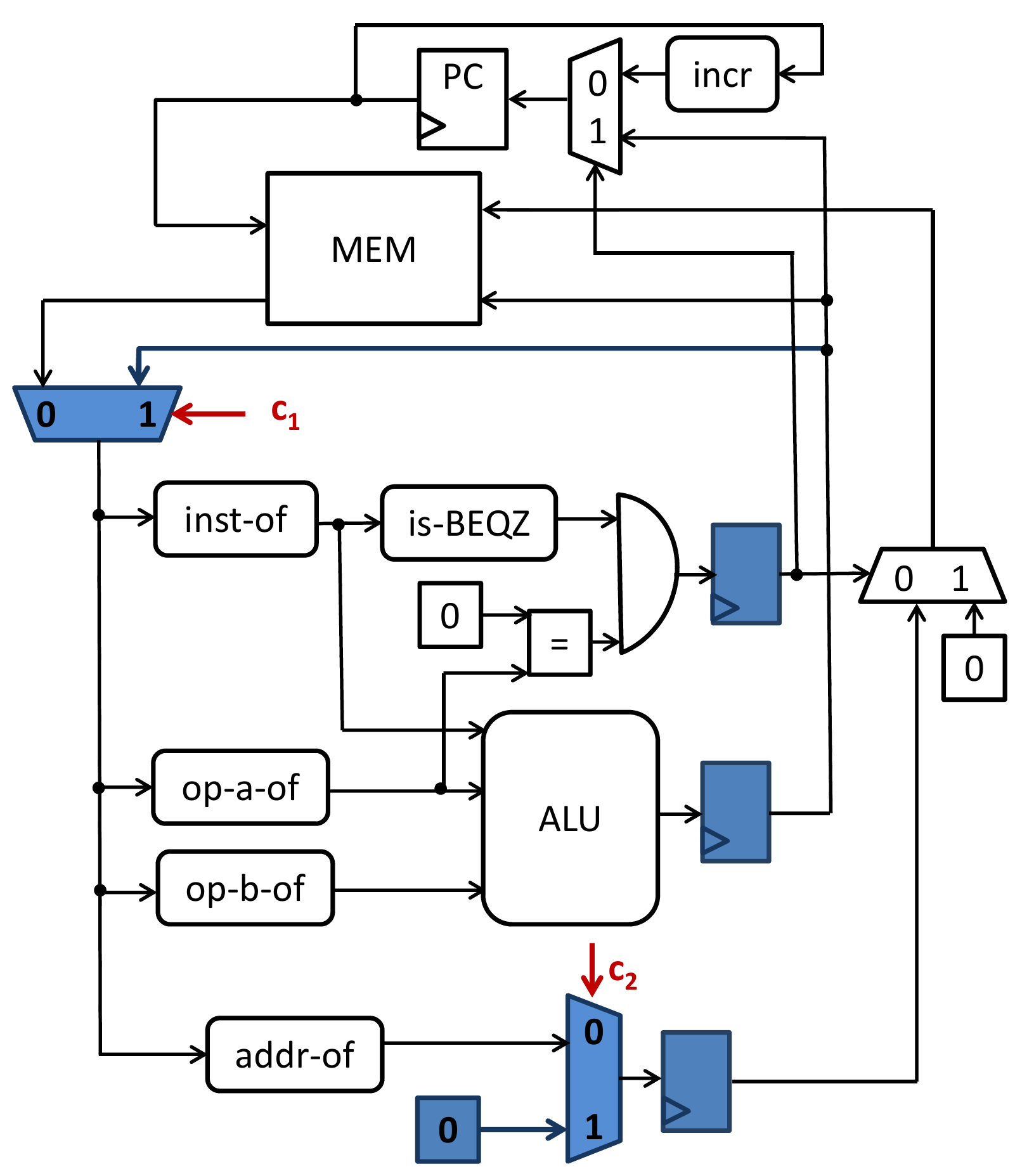}
    \vspace{-2ex}
  \caption{A simple microprocessor with a 2-stage pipeline.}
  \label{fig:simple_processor}
  \end{center}
  \vspace{-6ex}
\end{figure}

We used our tool on several small examples
and also tried one non-trivial example which we explain in more
detail. In Fig.~\ref{fig:simple_processor} we show a simple
(fictitious) microprocessor with a 2-stage pipeline. \textsf{MEM}
represents the main memory. We assume that the value at address 0 is
hardwired to 0. I.e., reading from address 0 always yields value 0.
The blocks \textsf{inst-of}, \textsf{op-a-of}, \textsf{op-b-of}, and
\textsf{addr-of} represent combinational functions that decode a
memory word. The block \textsf{incr}  increments the program counter
(\textsf{PC}). The block \textsf{is-BEQZ} is a predicate that checks
whether an instruction is a branch instruction. The design has two
pipeline-related control signals for which we would like to
synthesize an implementation. Signal $c_1$ causes a value
 in the pipeline to be forwarded and signal $c_2$ squashes the instruction that is currently
decoded and executed in the first pipeline stage. This might be
necessary due to speculative execution based on a ``branch-not-taken
assumption''.
The implementation of these control signals is not as
simple as it might seem at first glance. For example, the seemingly
trivial solution of setting $c_1=\ltrue$ whenever \textsf{PC} equals
the address register is not correct. For example, if $\textsf{PC}=0$,
forwarding should not be done.\footnote{We actually made this mistake
while trying to create and model-check a manual implementation for
the control signals, and it took some time to locate and understand
the problem.} By taking out the blue parts in
Fig.~\ref{fig:simple_processor} we obtain the non-pipelined reference
implementation which we used to formulate a Burch-Dill-style
equivalence criterion~\cite{Burch94}. The resulting formula was used
as a specification for synthesis.

\begin{table}[t]
\caption{\emph{Experimental results.} \textup{Columns: (1) Name; (2)
Number of control signals; (3) Total synthesis time including
checking the results; (4) Number of leaves with theory-tautology
clauses that define a new non-colorable literal (Number of such
leaves at the start of the cleaning procedure + Number of leaves
introduced (and subsequently removed) by the procedure); (5) Number
of leaves to be split because they contain literals from more than
one partition. (Number after ``$/$'' is total number of leaves in
proof at beginning of split procedure; (6) Time to reorder the proof
to be \lfirst; (7) Number of nodes in proof from \verit~/ Size of the
transformed proof for interpolation (local sub-trees have been
converted to leaves).}}
\label{tab:experimental_results} %
\vspace{-2ex}
\centering
\begin{tabular}{p{2.3em} p{1em} p{3.0em} p{3.2em} p{3.3em} p{3.8em} p{4.2em}}
\toprule
Name & Ctrl &
time~[s] &
\# leaves to clean &
\# leaves to split &
Reorder-Time~[ms] &
Proof size \\
\midrule
const  & 2 &   0.6 &  0      &  0~/~6     &    42 &    19~/~1    \\
illu02 & 2 &   1.1 &  1      &  1~/~65    &    83 &   205~/~12   \\
illu03 & 3 &   5.0 &  8      &  8~/~138   &   487 &   467~/~22   \\
illu04 & 4 &   8.0 &  3      &  3~/~242   &   532 &   951~/~75   \\
illu05 & 5 &  12.8 & 10      & 10~/~413   &   589 &  1588~/~78   \\
illu06 & 6 & 237.0 &  9      &  9~/~1093  &  1820 &  4691~/~370  \\
illu07 & 7 & 150.0 & 14      & 14~/~1443  &  2860 &  6824~/~555  \\
illu08 & 8 &1270.0 & 20      & 20~/~3450  &  4980 & 17524~/~1023  \\
pipe   & 1  &   1.6 & 6 + 6  &  3~/~70    &   129 &   285~/~22   \\
proc   & 2  &  28.1 & 3 + 3  &  61~/~1014 &  1770 &  5221~/~1042 \\
\bottomrule
\end{tabular}

\vspace{-5ex}
\end{table}


Table~\ref{tab:experimental_results} summarizes our experimental
results. The benchmark ``const'' is a simple example with 2 control
signals that allows for constants as valid solutions. ``illu02'' is
the example presented in Section~\ref{sec:illus}; ``illu03'' to
``illu08'' are scaled-up versions of ``illu02'', with increased
numbers of inputs and control signals. ``pipe'' is the simple
pipeline example that was used in~\cite{Hoffer11}. ``proc'' is the
pipelined processor shown in Fig.~\ref{fig:simple_processor} and
described above. All experiments were performed on an Intel Nehalem
CPU with $3.4$ GHz.

Note that using our new method we have reduced the synthesis time of
``pipe'' from 14 hours~\cite{Hoffer11} to $1.6$ seconds. As a second
comparison, we tried to reduce the (quantified) input formula of
``proc'' to a QBF problem (using the transformations outlined
in~\cite{Hoffer11}) and run \depqbf~\cite{Lonsin10} on it. After
approximately one hour, \depqbf exhausted all $192$ GB of main memory
and terminated without a result.


\section{Conclusion}
\label{sec:conclusion}

Hofferek and Bloem~\cite{Hoffer11} have shown that uninterpreted
functions are an efficient way to abstract away unnecessary details
in controller synthesis problems. By using interpolation in
$\theory_U$, we avoid the costly reduction to propositional logic,
thus unleashing the full potential of the approach presented
in~\cite{Hoffer11}. Furthermore, by introducing the concept of
$n$-interpolation, we also avoid the iterative construction which
requires several calls to the SMT solver and back-substitution.
The $n$-interpolation approach improves synthesis times by several
orders of magnitude, compared to previous methods~\cite{Hoffer11},
rendering it applicable to real-world problems, such as pipelined
microprocessors. We have also shown that a naive transformation to
QBF is not a feasible option.



\bibliographystyle{abbrv}
\bibliography{biblio}

\iftoggle{cameraready}{ } {

\appendix
\subsection{Why local-first?} \label{why-local-first}

\begin{example}
  Lets consider the formula $\forall a, b \; \exists c_1, c_2 \; \forall l\; . \; \phiValid(a,b,c_1,c_2,l)$,
where
$$
\begin{array}{@{}l@{\;}l@{}}
  \phiValid(a,b,c_1,c_2,l) := &
   (c_1 \land \neg c_2 \land \neg a) \lor
  (\neg c_1 \land c_2 \land \neg b) \lor \\
  &
  (c_1 \land c_2 \land ( (\neg l \land a )\lor (l \land b) ) ),\\
\end{array}
$$
and all variables are Boolean. We will compute witness functions for
$c_1$ and $c_2$  in terms of $a$ and $b$. After instantiating for all
values of $c_1$ and $c_2$, and negating the instantiated formulas, we
obtain
$$
\begin{array}{@{}l@{}}
\begin{array}{@{}l@{\qquad}l@{}}
  \phiUnsat_{FF} := \ltrue &   \phi_{FT} := b \\
  \phiUnsat_{TF} := a &
  \phiUnsat_{TT} := (l \lor \neg a )\land (\neg l \lor \neg b)\\
\end{array}\\
  \phiUnsat := \phiUnsat_{FF} \land \phiUnsat_{FT}  \land \phiUnsat_{TF} \land \phiUnsat_{TT}.
\end{array}
$$
The global variables are $G=\{a,b\}$. Since $l$ appears only in
$\phiUnsat_{TT}$, we decided not to rename it. Lets suppose the
following is the proof of unsatisfiability of $\phiUnsat$ produced by
an SMT solver.
$$
\inference{\Res}{
  \inference{\mathllap\Res}{
     a \quad  l \lor \neg a}{ l}{}\quad
  \inference{\Res}{ b \quad
     \neg l \lor \neg b}{ \neg l}{}
}{ \emptyset}{}
$$
In the above proof, $l$ is used as a pivot in the very last proof
step. Thus, the proof violates the local-first property. If we
compute an interpolant between $\phiUnsat_{FF} \land \phiUnsat_{FT}$
and $\phiUnsat_{TF} \land \phiUnsat_{TT}$ by applying the original
interpolation rules on the proof then we obtain the following
annotated proof.
$$
\inference{\Res}{
  \inference{\mathllap\Res}{
     a[\ltrue] \quad  l \lor \neg a[\ltrue]}
  { l[\ltrue]}{}\quad
  \inference{\Res}{
     b[\lfalse] \quad  \neg l \lor \neg b[\ltrue]}
  { \neg l[b]}{}
}{ \emptyset[b]}{}
$$
Similarly, if we compute an interpolant between $\phiUnsat_{FF} \land
\phiUnsat_{TF}$ and $\phiUnsat_{FT} \land \phiUnsat_{TT}$ by applying
the original interpolation rules on the proof then we obtain the
following annotated proof.
$$
\inference{\Res}{
  \inference{\mathllap\Res}{
     a[\lfalse] \quad  l \lor \neg a[\ltrue]}
  { l[a]}{}\quad
  \inference{\Res}{ b[\ltrue] \quad
     \neg l \lor \neg b[\ltrue]}
  { \neg l[\ltrue]}{}
}{ \emptyset[a]}{}
$$
From the above annotations, we learn $I_{c_1} = a$ and $I_{c_1} = b$,
which are not valid witness functions if they are used together
because $\phiValid(a,b,a,b,l)$ is not a valid formula. Lets consider
the following unsatisfiability proof, which satisfies the local-first
property.
$$
\begin{array}{@{}l@{}}
    \inference{\Res}{ l \lor \neg a \quad
       \neg l \lor \neg b}
    { \neg a \lor \neg b}{}\vspace{-10pt}\\
    \hspace{5ex}
    \inference{\Res}{\hspace{17ex} a}{
       \neg b}{}\vspace{-10pt}\\
    \hspace{13ex}\inference{\Res}{\hspace{13ex} b}
    { \emptyset}{}
\end{array}
$$
If we annotate this proof with our annotation rules, we obtain
$$
\begin{array}{@{}l@{}}
    \inference{\Res}{ \lor \neg a[(\ltrue,\ltrue)] \quad
       \neg l \lor \neg b[(\ltrue,\ltrue)]}
    { \neg a \lor \neg b[(\ltrue,\ltrue)]}{}\vspace{-10pt}\\
    \hspace{9ex}\inference{\Res}{\hspace{20ex} a[(\ltrue,\lfalse)]}{
      \quad\quad \neg b[(\ltrue,a)]
    }{}\vspace{-9pt}\\
    \hspace{20ex}\inference{\Res}{\hspace{15ex} b[(\lfalse,\ltrue)]}{
       \emptyset[(b,\neg b \lor a)]}{}.
\end{array}
$$
From the above annotations we learn $I_{c_1} = b$ and $I_{c_1} = \neg
b \lor a $, which are valid witness functions.
\end{example}


\subsection{Proof of Theorem~\ref{thm:intpol-is-witness}}
\label{proof:intpol-is-witness}

\begin{proof}
Since $\os_\lfalse$ and  $\os_\ltrue$ do not appear in both
$\neg\phiValid(\is,\lfalse,\os_\lfalse)$ and $\neg
\phiValid(\is,\ltrue,\os_\ltrue)$, only $\is$ can appear in the
interpolant between the formulas. Let $I(\is)$ be the interpolant
computed via our interpolation procedure. Due to the definition of
interpolation, $ \neg\phiValid(\is,\lfalse,\os_\lfalse) \limplies
I(\is) \text{ and } \neg\phiValid(\is,\ltrue,\os_\ltrue) \limplies
\neg I(\is). $ After reversing the implications, we obtain $ \neg
I(\is) \limplies \phiValid(\is,\lfalse,\os_\lfalse) \text{ and }
I(\is) \limplies \phiValid(\is,\ltrue,\os_\ltrue). $ After
reintroducing the quantifiers for $\os$, we obtain $ \neg I(\is)
\limplies \forall \os.\phiValid(\is,\lfalse,\os) \text{ and } I(\is)
\limplies \forall \os.\phiValid(\is,\ltrue,\os). $ Therefore, $
\forall \is. \forall \os.\phiValid(\is,I(\is),\os). $ Hence, $I(\is)$
is a witness function for $c$.
\end{proof}


\subsection{Proof of Theorem~\ref{thm:n-intpol-are-witnesses}}
\label{proof:n-intpol-are-witnesses}
\begin{proof}
Since for each $w$, $\os_w$ only appear in $\phiUnsat_w$, they cannot
be in $G$ and only $\is$ are in $G$. Let $\vecI(\is) = ( I_1(\is),
\dots, I_n(\is) )$ be an $n$-interpolant w.r.t.\ $\phiUnsat_w$. Due
to the definition of $n$-interpolation, for each $w \in \booleans^n$,
$$
\phiUnsat_w \limplies \Lor \vecI(\is) \xor w.
$$
After reversing the implications and expanding the definition of
$\phiUnsat_w$, we obtain
$$
\Land \neg\vecI(\is)\xor w \limplies \phiValid(\is,w,\os_w).
$$
After reintroducing the quantifiers for $\os$, we obtain
$$
\Land \neg\vecI(\is)\xor w \limplies \forall \os. \phiValid(\is,w,\os).
$$
Therefore, $\forall \is.\; w = \vecI(\is) \limplies \forall
\os.\phiValid(\is, w,\os).$ Therefore,
$$\forall \is. \forall \os.\phiValid(\is,\vecI(\is),\os).$$
Hence for each
$j\in \betweenOf{1}{n}$, $I_j(\is)$ is a witness function for $c_i$.
\end{proof}


\subsection{Proof of Theorem~\ref{thm:annotations-are-interpolants} }
\label{proof:annotations-are-interpolants}

\begin{proof} We will prove the theorem using induction over the proof
structure.
All annotations must satisfy the conditions of $n$-partial interpolant.
Note that $\Lor w \xor w = \lfalse$ and
if $w' \neq w$ then $\Lor w' \xor w = \ltrue$.

\underline{Base cases:}\\
\underline{\mHyp:} Let $w'=w$. Since $C \in \phiUnsat_w$, $C|_{w'} =
C $. Therefore, $ C|_{w'} \lor \Lor w \xor w' = C. $ Since $C \in
\phiUnsat_w$, $\phiUnsat_{w'} \limplies C$. Therefore, $
\phiUnsat_{w'} \limplies C|_{w'} \lor \Lor w \xor {w'}. $ Now, let
$w' \neq w$. Therefore, $ C|_{w'} \lor \Lor w \xor w' = \ltrue. $
Therefore, $ \phiUnsat_{w'} \limplies C|_{w'} \lor \Lor w \xor w'. $
$w \preceq G$. \\
\underline{\mAxi:} Let $w'=w$. Since $C \preceq \phiUnsat_w$,
$C|_{w'} = C $. Therefore, $ C|_w \lor \Lor w \xor w = C. $ Since $C$
is a theory tautology clause, $ \phiUnsat_w \limplies C|_w \lor \Lor
w \xor w. $ For $w' \neq w$, same as previous rule. $w \preceq G$.

\underline{Inductive cases:} We assume that the annotations of the
premises are $n$-partial
interpolant.\\
\underline{\mRes:}
Let $w'=w$. Since
$$
\phiUnsat_{w'} \limplies a \lor C  \lor \Lor w \xor {w'}
\text{ and }
\phiUnsat_{w'} \limplies \neg a \lor D  \lor \Lor w \xor {w'},
$$
we can easily show $ \phiUnsat_{w'} \limplies C \lor D  \lor \Lor w
\xor w'. $ For $w' \neq w$, same as previous rule.
$w \preceq G$.\\
\underline{\mResG:}
Due to the definition of $n$-partial interpolants, for each $w \in \booleans^n$
$$
\phiUnsat_{w} \limplies a \lor C|_w  \lor \Lor \vecI^C \xor w
\text{ and }
\phiUnsat_{w} \limplies \neg a \lor D|_w  \lor \Lor \vecI^D \xor w.
$$
After taking conjunction of the two implications, we obtain
$$
\phiUnsat_{w} \limplies (a \lor C|_w  \lor \Lor \vecI^C \xor w )  \land
(\neg a \lor D|_w  \lor \Lor \vecI^D \xor w).
$$
After expanding the definition of or-of-xor of vectors and
moving $a$ and $\neg a$ inside the vector disjunction, we obtain
$$
\phiUnsat_{w} \limplies ( C|_w  \lor \Lor_{j=0}^n ( a \lor I^C_j \xor w_j)  )
\land
( D|_w  \lor \Lor_{j=0}^n (\neg a \lor I^D_j \xor w_j) ).
$$
Using $(a \lor b) \land ( c \lor d) \limplies (a\lor c) \lor (b\land d)$, we obtain
$$
\phiUnsat_{w} \limplies (C \lor D)|_w \lor
( \Lor_{j=0}^n ( a \lor I^C_j \xor w_j)  \land
 \Lor_{j=0}^n (\neg a \lor I^D_j \xor w_j) ).
$$
After moving out the disjunctions, we obtain
$$
\phiUnsat_{w} \limplies (C \lor D)|_w \lor
\Lor_{j=0}^n  (( a \lor I^C_j \xor w_j)  \land  (\neg a \lor I^D_j \xor w_j) ).
$$
After moving out the $\xor$ operator, we obtain
$$
\phiUnsat_{w} \limplies (C \lor D)|_w \lor
\Lor_{j=0}^n  (( a \lor I^C_j )  \land  (\neg a \lor I^D_j) )\xor w_j.
$$
Since $\{a,\vecI^C,\vecI^D\} \preceq G$, symbols in the annotation of
the conclusion are within $G$.
\end{proof}



}

\end{document}